\documentclass[prc,aps,superscriptaddress,twocolumn,floatfix,showpacs,showkeys,altaffilletter]{revtex4-1}

\newcommand{\nuc}[2]{$^{#1}$#2}

\usepackage{graphicx}                  % use to insert eps images into document
\usepackage{pdfpages}                  % use for inserting PDF's into the document
\usepackage[rightcaption]{sidecap}
\usepackage{epsfig}
\usepackage{bm}
\usepackage{verbatim}   
\usepackage{color}         
\usepackage[percent]{overpic}
\usepackage{multirow}
\usepackage{longtable}
\usepackage{tablefootnote}
\usepackage{dcolumn}
\usepackage{setspace}
\usepackage{array}
\usepackage{epstopdf}
\usepackage{hyperref}
\hypersetup{
     colorlinks   = true,
     citecolor    = blue,
     urlcolor     = blue,
     linkcolor   = blue,
}

\begin{document}

%%%%%%%%%%%%%%%%%%%%%%%%%%%%%%%%%%%%%%%%%
%%%% Title
%%%%%%%%%%%%%%%%%%%%%%%%%%%%%%%%%%%%%%%%%

\title{\textbf{$\gamma$}-soft \nuc{146}{Ba} and the role of non-axial shapes at \textbf{\textit{N}} $\sim$ 90}

%%%%%%%%%%%%%%%%%%%%%%%%%%%%%%%%%%%%%%%%%
%%%% Author list
%%%%%%%%%%%%%%%%%%%%%%%%%%%%%%%%%%%%%%%%%

\renewcommand{\thefootnote}{\fnsymbol{footnote}}

\author{A.~J.~Mitchell}
\altaffiliation[Present address: ]{Department of Nuclear Physics, Australian National University, Canberra, ACT 2601, Australia}
\email[Email: ]{aj.mitchell@anu.edu.au}
\affiliation{Department of Physics and Applied Physics, University of Massachusetts Lowell, Lowell, Massachusetts 01854}

\author{C.~J.~Lister}
\affiliation{Department of Physics and Applied Physics, University of Massachusetts Lowell, Lowell, Massachusetts 01854}

\author{E.~A.~McCutchan}
\affiliation{National Nuclear Data Center, Brookhaven National Laboratory, Upton, New York 11973}

\author{M.~Albers}
\altaffiliation[Present address: ]{Ernst $\&$ Young GmbH, Wirtschaftspruefungsgesellschaft, Mergenthalerallee 3-5, D-65760 Eschborn, Germany}
\affiliation{Physics Division, Argonne National Laboratory, Argonne, Illinois 60439}

\author{A.~D.~Ayangeakaa}
\affiliation{Physics Division, Argonne National Laboratory, Argonne, Illinois 60439}

\author{P.~F.~Bertone}
\altaffiliation[Present address: ]{Marshall Space Flight Center, Huntsville, Alabama 35812}
\affiliation{Physics Division, Argonne National Laboratory, Argonne, Illinois 60439}

\author{M.~P.~Carpenter}
\affiliation{Physics Division, Argonne National Laboratory, Argonne, Illinois 60439}

\author{C.~J.~Chiara}
\altaffiliation[Present address: ]{U.S. Army Research Laboratory, Adelphi, Maryland 20783}
\affiliation{Physics Division, Argonne National Laboratory, Argonne, Illinois 60439}
\affiliation{Department of Chemistry and Biochemistry, University of Maryland, College Park, Maryland 20742}

\author{P.~Chowdhury}
\affiliation{Department of Physics and Applied Physics, University of Massachusetts Lowell, Lowell, Massachusetts 01854}

\author{J.~A.~Clark}
\affiliation{Physics Division, Argonne National Laboratory, Argonne, Illinois 60439}

\author{P.~Copp}
\affiliation{Department of Physics and Applied Physics, University of Massachusetts Lowell, Lowell, Massachusetts 01854}

\author{H.~M.~David}
\altaffiliation[Present address: ]{GSI Helmholtzzentrum f\"{u}r Schwerionenforschung GmbH, D-64291 Darmstadt, Germany}
\affiliation{Physics Division, Argonne National Laboratory, Argonne, Illinois 60439}

\author{A.~Y.~Deo}
\altaffiliation[Present address: ]{Department of Physics, Indian Institute of Technology Roorkee, Roorkee 247 667, India}
\affiliation{Department of Physics and Applied Physics, University of Massachusetts Lowell, Lowell, Massachusetts 01854}

\author{B.~DiGiovine}
\affiliation{Physics Division, Argonne National Laboratory, Argonne, Illinois 60439}

\author{N.~D'Olympia}
\altaffiliation[Present address: ]{Passport Systems Inc., 70 Treble Cove Road, 1st Floor, Billerica, Massachusetts 01862}
\affiliation{Department of Physics and Applied Physics, University of Massachusetts Lowell, Lowell, Massachusetts 01854}

\author{R.~Dungan}
\affiliation{Physics Department, Florida State University, Tallahassee, Florida 32306}

\author{R.~D.~Harding}
\altaffiliation[Present address: ]{Department of Physics, University of York, Heslington, York, YO10 5DD, UK}
\affiliation{Department of Physics and Applied Physics, University of Massachusetts Lowell, Lowell, Massachusetts 01854}
\affiliation{Department of Physics, University of Surrey, Guildford GU2 7XH, UK}

\author{J.~Harker}
\affiliation{Physics Division, Argonne National Laboratory, Argonne, Illinois 60439}
\affiliation{Department of Chemistry and Biochemistry, University of Maryland, College Park, Maryland 20742}

\author{S.~S.~Hota}
\altaffiliation[Present address: ]{Department of Nuclear Physics, Australian National University, Canberra, ACT 2601, Australia}
\affiliation{Department of Physics and Applied Physics, University of Massachusetts Lowell, Lowell, Massachusetts 01854}

\author{R.~V.~F.~Janssens}
\affiliation{Physics Division, Argonne National Laboratory, Argonne, Illinois 60439}

\author{F.~G.~Kondev}
\affiliation{Nuclear Engineering Division, Argonne National Laboratory, Argonne, Illinois 60439}

\author{S.~H.~Liu}
\altaffiliation[Present address: ]{West Physics, 3825 Paces Walk SE, Suite 250, Atlanta, Georgia 30339}
\affiliation{Department of Chemistry, University of Kentucky, Lexington, Kentucky 40506}
\affiliation{Department of Physics and Astronomy, University of Kentucky, Lexington, Kentucky 40506}

\author{A.~V.~Ramayya}
\affiliation{Physics Department, Vanderbilt University, Nashville, Tennessee 37235}

\author{J.~Rissanen}
\altaffiliation[Present address: ]{Fennovoima Oy, Salmisaarenaukio 1, 00180 Helsinki, Finland}
\affiliation{Lawrence Berkeley National Laboratory, Berkeley, California 94720}

\author{G.~Savard}
\affiliation{Physics Division, Argonne National Laboratory, Argonne, Illinois 60439}
\affiliation{Department of Physics, University of Chicago, Chicago, Illinois 60637}

\author{D.~Seweryniak}
\affiliation{Physics Division, Argonne National Laboratory, Argonne, Illinois 60439}

\author{R.~Shearman}
\altaffiliation[Present address: ]{National Physical Laboratory, Hampton Road, Teddington, Middlesex TW11 0LW, UK}
\affiliation{Department of Physics and Applied Physics, University of Massachusetts Lowell, Lowell, Massachusetts 01854}
\affiliation{Department of Physics, University of Surrey, Guildford GU2 7XH, UK}

\author{A.~A.~Sonzogni}
\affiliation{National Nuclear Data Center, Brookhaven National Laboratory, Upton, New York 11973}

\author{S.~L.~Tabor}
\affiliation{Physics Department, Florida State University, Tallahassee, Florida 32306}

\author{W.~B.~Walters}
\affiliation{Department of Chemistry and Biochemistry, University of Maryland, College Park, Maryland 20742}

\author{E.~Wang}
\affiliation{Physics Department, Vanderbilt University, Nashville, Tennessee 37235}

\author{S.~Zhu}
\affiliation{Physics Division, Argonne National Laboratory, Argonne, Illinois 60439}

\date{\today}

%%%%%%%%%%%%%%%%%%%%%%%%%%%%%%%%%%%%%%%%%
%%%% Abstract
%%%%%%%%%%%%%%%%%%%%%%%%%%%%%%%%%%%%%%%%%

\begin{abstract}

Low-spin states in the neutron-rich, $N = 90$ nuclide \nuc{146}{Ba} were populated following $\beta$-decay of \nuc{146}{Cs}, with the goal of clarifying the development of deformation in Ba isotopes through delineation of their non-yrast structures. Fission fragments of \nuc{146}{Cs} were extracted from a 1.7-Ci \nuc{252}{Cf} source and mass-selected using the CARIBU facility. Low-energy ions were deposited at the center of a box of thin $\beta$ detectors, surrounded by a high-efficiency HPGe array. The new \nuc{146}{Ba} decay scheme now contains 31 excited levels extending up to $\sim$2.5~MeV excitation energy, double what was previously known. These data are compared to predictions from the Interacting Boson Approximation (IBA) model. It appears that the abrupt shape change found at $N = 90$ in Sm and Gd is much more gradual in Ba and Ce, due to an enhanced role of the $\gamma$ degree of freedom. 

\end{abstract}

\pacs{23.40.-s, 21.60.Fw, 23.20.Lv}
\keywords{$\beta$~decay; $\gamma$-ray spectroscopy; $E_{\gamma}$, $I_{\gamma}$; $N = 90$ shape transition; Interacting Boson Approximation}

\maketitle

%%%%%%%%%%%%%%%%%%%%%%%%%%%%%%%%%%%%%%%%%
%%%% Introduction
%%%%%%%%%%%%%%%%%%%%%%%%%%%%%%%%%%%%%%%%%

\section{INTRODUCTION}\label{sec:introduction}

The transition from spherical, shell-model-like behavior, to deformed collective motion has always been interesting, yet controversial, in nuclear structure. Although models exist for each extreme \cite{Reference1,Reference2}, the actual transition from one limit to the other remains confused and lacks a ubiquitous description. Stable isotopes of rare-earth elements near $Z = 64$ with $N = 90$ (e.g. \nuc{156}{Dy} ($Z = 66$) \cite{Reference3}, \nuc{154}{Gd} ($Z = 64$) \cite{Reference4}, \nuc{152}{Sm} ($Z = 62$) \cite{Reference5}, and \nuc{150}{Nd} ($Z = 60$) \cite{Reference6}) exhibit remarkable similarities in the excitation energies of ground-state bands and excited $J^{\pi}$~=~0$^+$ and $J^{\pi}$ = 2$^+$ sequences. The abrupt onset of deformation has received particularly intense scrutiny with general discussions often framed in terms of a phase transition \cite{Reference7,Reference8,Reference9}; in this case, a specific type of phase transition encapsulated by the X(5) model \cite{Reference10,Reference11,Reference12,Reference13}. However, such an approach is not fully supported by all available experimental data and more generalized shape-coexistence models have been proposed \cite{Reference14,Reference15,Reference16}. 

A way to clarify this issue is to widen the scope of investigation to both heavier and lighter nuclei. In a general sense, the behavior of transitional nuclei is expected to follow the number of valence particles, as predicted in the $N_pN_n$ scheme of Casten \cite{Reference17}. In practice, the underlying fermionic structure appears to be important, with residual interactions between protons and neutrons in specific orbits playing a key role in `tipping' the nuclear shape from spherical to deformed \cite{Reference18,Reference19}. In this way, the $N = 90$ border between shapes retains its significance, although the sharpness of the transition becomes more muted. The nuclei which have been most extensively studied are all stable, but it is relevant to enquire about how the transitional structures evolve as one progresses up to \nuc{158}{Er} ($Z = 68$), and \nuc{160}{Yb} ($Z = 70$) or down to \nuc{148}{Ce} ($Z = 58$), and \nuc{146}{Ba} ($Z = 56$). The lighter nuclei in this sequence are quite neutron-rich and cannot be accessed by fusion-evaporation reactions, and so fission-fragment spectroscopy and $\beta$~decay are the appropriate probes. 

Nuclei in this region are also expected to exhibit strong octupole correlations \cite{Reference20}. Polarization of spin-orbit partners appears to quench the $Z = 64$ sub-shell closure, resulting in strong couplings between $\Delta J = \Delta L = 3$ nucleon orbitals ($\pi d_{5/2} - \pi h_{11/2}$ and $\nu f_{7/2} - \nu i_{13/2}$). The abrupt onset of octupole collectivity in Gd, Sm, and Nd is observed between $N~=~88$ and $N~=~90$. The Ba isotopes undergo a smoother transition between $N = 86$ and $N = 88$, two neutrons earlier than expected from the behavior found in the $Z = 60$ to $Z = 64$ range \cite{Reference21}. Prompt-fission spectroscopy of \nuc{146}{Ba} (for example \cite{Reference22,Reference23,Reference24}) has identified the ground-state and negative-parity bands to moderate spins. Octupole deformation in \nuc{146}{Ba} has been discussed  \cite{Reference25,Reference26,Reference27}, with the suggestion that these effects are weak and disappear at medium to high spins \cite{Reference24}. Although the yrast states in \nuc{146}{Ba} are established, there is limited information pertaining to the non-yrast, low-spin levels. This paper reports on the $\beta$-decay of \nuc{146}{Cs}, with focus on identifying and quantifying properties of the important low-spin, non-yrast states in \nuc{146}{Ba} which inform this discussion.  

%%%%%%%%%%%%%%%%%%%%%%%%%%%%%%%%%%%%%%%%%
%%%% Experimental set up
%%%%%%%%%%%%%%%%%%%%%%%%%%%%%%%%%%%%%%%%%

\section{EXPERIMENTAL METHOD}\label{sec:experiment}

The data presented here were obtained at the CAlifornium Rare Ion Breeder Upgrade ({CARIBU} \cite{Reference28}) facility at Argonne National Laboratory. Spontaneous fission fragments extracted from a 1.7-Ci \nuc{252}{Cf} source were thermalized in a gas catcher, in which interactions with high-purity He gas and with RF and DC fields combine to result in a low-emittance beam. An isotopically-pure beam of singly-charged \nuc{146}{Cs} nuclei was selected by the isobar separator. The beam was cooled to $\sim$2~keV and bunched before delivery to the low-energy experimental area. Approximately 300~ions/s were delivered to the new decay-spectroscopy station, where they were implanted on an aluminum foil located at the center of an array of $\gamma$-ray and $\beta$-particle detectors. 

The CARIBU decay station consists of the SATURN (Scintillator And Tape Using Radioactive Nuclei) system coupled to the X-Array, a highly-efficient array of five High-Purity Ge (HPGe) clover detectors. The measurement described in this article utilized the `Mark-\text{I}' detector chamber with the `paddle' scintillator arrangement (see Ref. \cite{Reference29} for a description of the experimental set-up). These data were measured as part of the commissioning for the new decay station. The catcher foil was located at the geometric center of four symmetrically-arranged plastic scintillator paddles, and replaced periodically to reduce the buildup of decay chain activity over time. Each paddle was positioned in front of a single clover in the vertical plane of the X-Array, offering large solid angle coverage. Output energy signals from each of the clover crystals and four scintillator preamps were fed directly into a digital data acquisition system. 

\begin{figure}[t!]
\includegraphics[width=8.5cm]{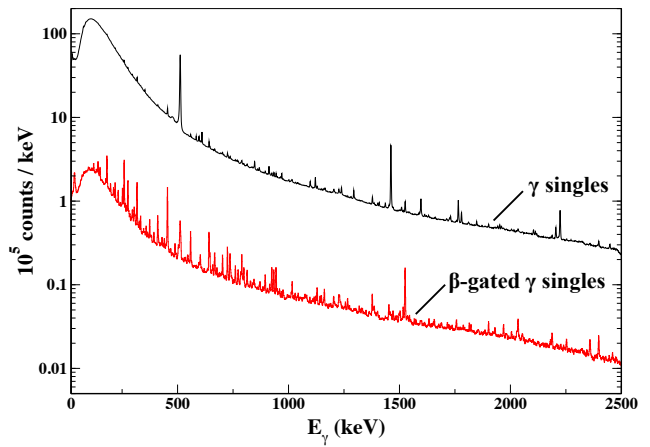}
\caption{\label{Figure1} [Color online] Total HPGe $\gamma$-ray energy spectra measured from $^{146}$Cs $\beta$~decay. The black upper spectrum is ungated $\gamma$-singles and the red (light gray) lower spectrum is $\beta$-gated $\gamma$-singles for the same data. By only selecting time-correlated $\beta$-$\gamma$ events, overall background is suppressed by up to two orders of magnitude and uncorrelated, room-background $\gamma$~rays are removed.}
\end{figure}

The energy spectrum of $\gamma$~rays detected by the X-Array is presented in Fig. \ref{Figure1}. The black upper spectrum represents the ungated $\gamma$-ray singles spectrum for all the data. The red (light gray) lower spectrum corresponds to Ge clover events detected in coincidence with an event in a scintillator paddle. Energy and efficiency calibrations of the X-Array were determined using standard \nuc{152}{Eu}, \nuc{182}{Ta}, and \nuc{243}{Am} sources. 

%%%%%%%%%%%%%%%%%%%%%%%%%%%%%%%%%%%%%%%%%
%%%% Results
%%%%%%%%%%%%%%%%%%%%%%%%%%%%%%%%%%%%%%%%%

\section{RESULTS}\label{sec:results}\label{sec:results}

%%%%%%%%%%%%%%%%%%%%%
\subsection{$\gamma$-ray identification}\label{subsec:gamma}
%%%%%%%%%%%%%%%%%%%%%

\begin{figure}[h!]
\includegraphics[width=8.5cm]{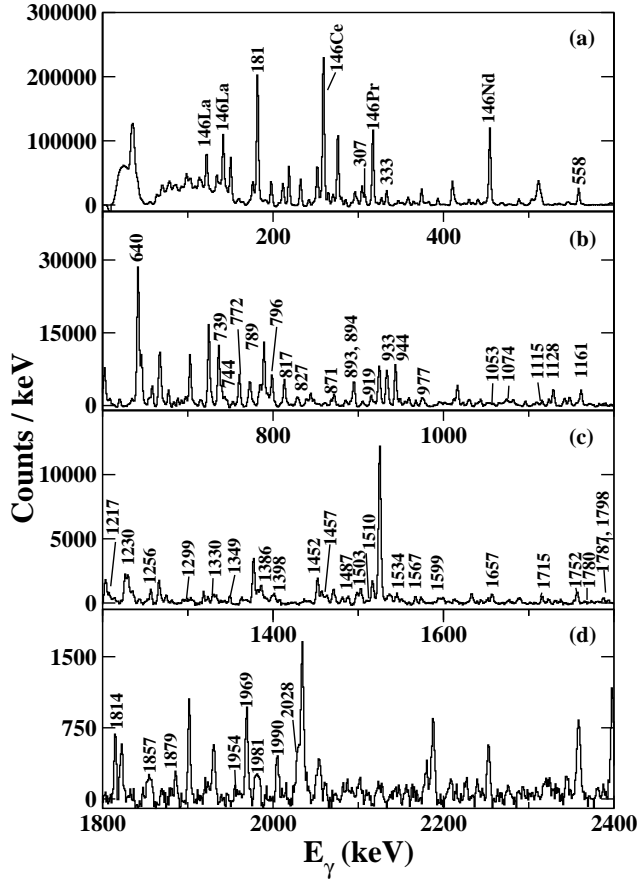}
\caption{\label{Figure2} Portions of the HPGe background-subtracted, $\beta$-gated $\gamma$-singles spectrum from (a) 0 to 600~keV, (b) 600 to 1200~keV, (c) 1200 to 1800~keV, and (d) 1800 to 2400~keV obtained from $\beta$~decay of \nuc{146}{Cs}. The identified $\gamma$~rays from $^{146}$Ba transitions are marked with their measured energies. Gamma rays from the strongest transitions in the long-lived activity of the $A = 146$ decay-chain sequence are also indicated. Unmarked $\gamma$~rays were identified as isobaric contaminants in the coincidence data.}
\end{figure}

For \nuc{146}{Cs}, the decay half-live, 0.321(2)~s, and $\beta$-delayed neutron branching ratio, 14.2(5)$\%$, are well known \cite{Reference30}. Gamma rays from each of the $A = 146$ isobars along the $\beta^-$ decay chain towards stability (\nuc{146}{Ba}$\rightarrow$\nuc{146}{La}$\rightarrow$\nuc{146}{Ce}$\rightarrow$\nuc{146}{Pr}$\rightarrow$\nuc{146}{Nd}) were recorded in the singles data. Those associated with the de-excitation of \nuc{146}{Ba} were identified using a combination of $\beta$-$\gamma$ and $\beta$-$\gamma$-$\gamma$ coincidence events. Contamination from long-lived activity was strongest in the even-even isobars, \nuc{146}{Ce} and \nuc{146}{Nd}. The odd-odd isobars, \nuc{146}{La} and \nuc{146}{Pr}, were highly fragmented and as such, their relative $\gamma$-ray intensities are weak. A small contribution from \nuc{145}{Ba}, from the $\beta$-delayed neutron emission of \nuc{146}{Cs}, was also detectable. The background-subtracted, $\beta$-gated $\gamma$-ray singles spectrum can be found in Fig. \ref{Figure2}. Gamma rays that have been identified as transitions in \nuc{146}{Ba} are labelled by their energies. The strongest transitions from the subsequent decay chain are also labelled. The full range of the energy spectrum in this measurement was $\sim$3~MeV. A short test has since been conducted with the range extended to $\sim$10~MeV; there was no evidence of any further strong, direct decays to the ground state beyond the range of the original experiment. 

\begin{center}
\begin{longtable}{@{\extracolsep{0.8cm}}cccc}

\caption{\label{Table1} Observed $\gamma$-ray transitions in \nuc{146}{Ba} placed in the level scheme of Fig. \ref{Figure5}. Relative intensities, $I_{\gamma}$, are normalized to the 181-keV $\gamma$~ray, taken as 100. For absolute intensity per 100 parent decays, the relative intensity should be multiplied by 0.42(5). The method for determining the normalization using the 141-keV $\gamma$~ray from \nuc{146}{La} \cite{Reference30} is described in the text (for reference, the relative intensity of this $\gamma$~ray is included in the table). Strong transitions were calculated from prompt $\gamma$-ray singles data; those marked $^{\dagger}$ are from coincidence data. Upper limits on $I_{\gamma}$ for transitions from new levels to the ground state that have not been observed, but may occur if $J^{\pi}_i\neq0$, are marked $^u$. Uncertainties are statistical and based on fitting approximations.} \\

\hline\hline \\[-0.3cm]
\multicolumn{1}{c}{$E_{\gamma}$} & 
\multicolumn{1}{c}{$I_{\gamma}$} & 
\multicolumn{1}{c}{$E_{\rm{initial}}$} & 
\multicolumn{1}{c}{$E_{\rm{final}}$} 	\\
\multicolumn{1}{c}{(keV) } & 
\multicolumn{1}{c}{} & 
\multicolumn{1}{c}{(keV)} & 
\multicolumn{1}{c}{(keV)} 	\\[0.1cm]
\hline  \\[-0.3cm]
\endfirsthead

\multicolumn{4}{c}%
{{\tablename\ \thetable{} -- continued}} \\
\hline\hline \\[-0.3cm]
\multicolumn{1}{c}{$E_{\gamma}$} & 
\multicolumn{1}{c}{$I_{\gamma}$} & 
\multicolumn{1}{c}{$E_{\rm{initial}}$} & 
\multicolumn{1}{c}{$E_{\rm{final}}$} 	\\
\multicolumn{1}{c}{(keV) } & 
\multicolumn{1}{c}{} & 
\multicolumn{1}{c}{(keV)} & 
\multicolumn{1}{c}{(keV)} 	\\[0.1cm]
\hline  \\[-0.3cm]
\endhead

 \\[-0.25cm]
\hline\hline  
\endfoot

 \\[-0.25cm]
\hline\hline 
\endlastfoot 

140.7(1)	&	41(2)						&		-	&	-		\\
181.3(1)	&	100(3)					&	181.1(1)	&	0.0		\\
307.3(1)	&	5.3(4)\text{$^{\dagger}$}	  	&	821.6(2)	&	513.9(2)	\\
332.9(1)	&	13(2)			        			& 	513.9(2)	&	181.1(1)	\\
558.1(1)	&	23(1) 					&	739.4(1)	&	181.1(1)	\\
639.9(1)	&	4.4(3)					&	821.6(2)	&	181.1(1)	\\
739.1(2)	&	5.1(6)					&	739.4(1)	&	0.0		\\
743.6(6)	&	2.7(6)					&	1566.2(2)	&	821.6(2)	\\
772.2(1)	&	5.2(6)					&	1511.7(2)	&	739.4(1)	\\
788.9(1)	&	0.50(6)\text{$^{\dagger}$}  	&	1529.1(1)	&	739.4(1)	\\
795.6(2)	&	2.0(7)					&	1309.5(3)		&	513.9(2)	\\
816.6(6)	&	1.2(5)					&	1638.2(3)		&	821.6(2)	\\
827.3(4)	&	2.2(9)					&	1566.2(2)	&	739.4(1)	\\
871.3(1)	&	3.4(6)					&	1052.4(3)		&	181.1(1)	\\
892.9(4)	&	1.24(11)\text{$^{\dagger}$}  	&	1632.6(2)	&	739.4(1)	\\
894.1(1)	&	0.21(4)\text{$^{\dagger}$}  	&	1714.9(2)	&	821.6(2)	\\
918.7(3)	&	1.4(5)					&	1657.3(2)	&	739.4(1)	\\
933.1(1)	&	5.2(4)\text{$^{\dagger}$}	&	1114.7(2)	&	181.1(1)	\\
943.6(2)	&	1.1(1)					&	1683.1(2)	&	739.4(1)	\\
976.7(1)	&	2.6(8)					&	1714.9(2)	&	739.4(1)	\\
1052.7(4)	&	1.5(7)					&	1566.2(2)	&	513.9(2)	\\
1073.5(2)	&	3.5(7)					&	1255.4(2)	&	181.1(1)		\\
1115.2(3)	&	2.9(5)					&	1114.7(2)	&	0.0			\\
1128.4(1)	&	2.7(2)\text{$^{\dagger}$}	&	1309.5(3)	&	181.1(1)		\\
1160.9(1)	&	1.2(1)\text{$^{\dagger}$}	&	1342.0(3)	&	181.1(1)		\\
1217(1)	&	0.6(5)					&	1397.8(2)	&	181.1(1)		\\
1229.5(2)	&	0.58(9)\text{$^{\dagger}$}		&	1410.8(3)	&	181.1(1)		\\
1256.1(3)	&	3.1(6)					&	1255.4(2)	&	0.0			\\
1299(1)	&	0.8(4)					&	2036.8(2)	&	739.4(1)	\\
1310(1)\text{$^{u}$}	&	\text{\textless}0.19		& 1309.5(3)   & 0.0\\ 
1330.4(2)	&	1.6(5)					&	1511.7(2)	&	181.1(1)		\\
1342(2)\text{$^{u}$}	&	\text{\textless}0.19		& 1342.0(3)  & 0.0 \\  
1348.9(3)	&	1.6(5)					&	1529.1(2)	&	181.1(1)		\\
1385.6(2)	&	4.3(7)					&	1566.2(2)	&	181.1(1)		\\
1397.8(4)	&	1.1(6)					&	1397.8(2)	&	0.0			\\
1412(1)\text{$^{u}$}	&	\text{\textless}0.20		& 1410.8(3)  & 0.0 \\
1451.8(1)	&	0.83(12)\text{$^{\dagger}$}	&	1632.6(2)	&	181.1(1)		\\
1457.0(2)	&	3.3(7)					&	1638.2(3)	&	181.1(1)		\\
1487.4(4)	&	2(1)						&	1668.5(2)	&	181.1(1)		\\
1502.5(2)	&	2.8(2)\text{$^{\dagger}$}	&	1683.1(1)	&	181.1(1)		\\
1510(1)	&	0.9(5)					&	1511.7(2)	&	0.0			\\
1529(1)\text{$^{u}$}	&	\text{\textless}0.21		& 1529.1(2)  & 0.0 \\
1533.7(5)	&	1.5(9)					&	1714.9(2)	&	181.1(1)		\\
1566.7(3)	&	2.6(5)					&	1566.2(17)	&	0.0			\\
1598.7(4)	&	2.3(6)					&	1780.0(2)	&	181.1(1)		\\
1633(1)\text{$^{u}$}	&	\text{\textless}0.23		& 1632.6(2)  & 0.0\\
1638(1)\text{$^{u}$}	&	\text{\textless}0.23		& 1638.2(3)  & 0.0\\
1656.6(4)	&	3.6(6)					&	1657.3(2)	&	0.0			\\
1669(1)\text{$^{u}$}	&	\text{\textless}0.23		& 1668.5(3)  & 0.0\\
1684(1)\text{$^{u}$}	&	\text{\textless}0.23		&  1683.1(2) & 0.0\\
1715.4(3)	&	2.7(6)					&	1714.9(2)	&	0.0			\\
1751.7(4)	&	0.79(14)\text{$^{\dagger}$}	&	1932.8(3)	&	181.1(1)		\\
1780.2(8)	&	0.9(6)					&	1780.0(2)	&	0.0			\\
1787.2(3)	&	2.3(6)					&	1968.5(2)	&	181.1(1)		\\
1798.3(4)	&	0.81(15)\text{$^{\dagger}$}	&	1979.4(3)	&	181.1(1)		\\
1814.4(2)	&	3.7(6)					&	1995.5(3)	&	181.1(1)		\\
1856.6(4)	&	1.5(2)\text{$^{\dagger}$}		&	2036.8(2)	&	181.1(1)		\\
1878.9(4)	&	1.0(2)\text{$^{\dagger}$}		&	2060.0(3)	&	181.1(1)		\\
1934(1)\text{$^{u}$}	&	\text{\textless}0.28		&  1932.8(3) & 0.0\\
1953.7(4)	&	1.1(2)\text{$^{\dagger}$}		&	2134.8(3)	&	181.1(1)		\\
1968.6(2)	&	7(1)						&	1968.5(2)	&	0.0			\\
1980(1)\text{$^{u}$}	&	\text{\textless}0.28		&  1979.4(3) & 0.0\\
1981.1(9)	&	2.2(9)					&	2162.2(3)	&	181.1(1)		\\
1990.2(5)	&	1.0(2)\text{$^{\dagger}$}		&	2171.3(3)	&	181.1(1)		\\
1996(1)\text{$^{u}$}	&	\text{\textless}0.28		&  1995.5(3) & 0.0\\
2027.8(4)	&	1.3(2)\text{$^{\dagger}$}		&	2208.9(3)	&	181.1(1)		\\
2037(1)\text{$^{u}$}	&	\text{\textless}0.29		& 2036.8(2)  & 0.0\\
2061(1)\text{$^{u}$}	&	\text{\textless}0.30		& 2060.0(3)  &0.0 \\
2136(1)\text{$^{u}$}	&	\text{\textless}0.31		& 2134.8(3)  & 0.0\\
2162(1)\text{$^{u}$}	&	\text{\textless}0.32		& 2162.2(3) &0.0\\
2172(1)\text{$^{u}$}	&	\text{\textless}0.32		& 2171.3(3)  &0.0 \\
2210(1)\text{$^{u}$}	&	\text{\textless}0.33		& 2208.9(3)  & 0.0\\

\end{longtable}
\end{center}

\begin{figure}[t!]
\includegraphics[width=8.5cm]{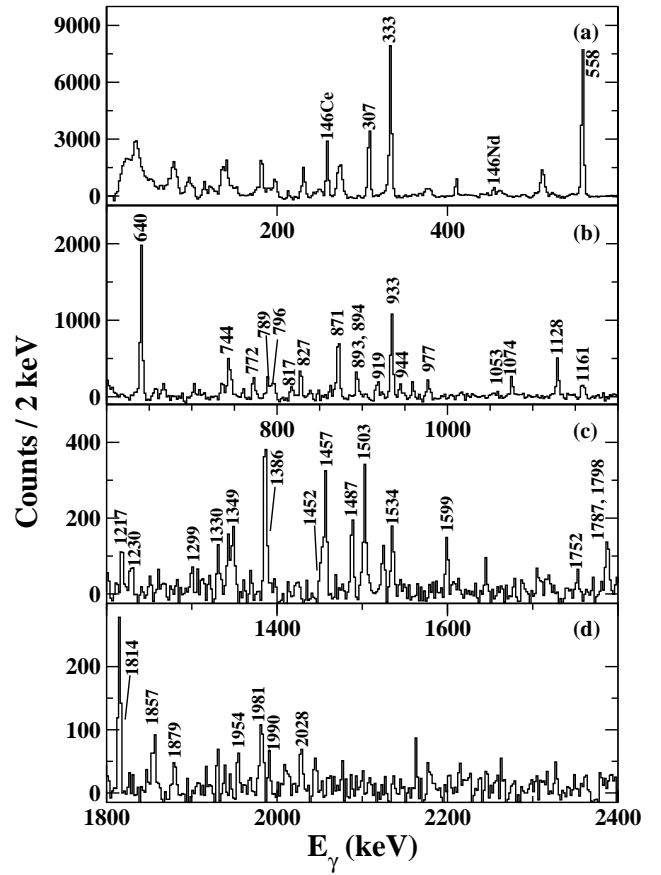}
\caption{\label{Figure3} Background-subtracted projection of the $\beta$-gated $\gamma$-$\gamma$ coincidence matrix, gated on the 181-keV, 2$^+_1$$\rightarrow$~0$^+_1$ transition from (a) 0 to 600~keV, (b) 600 to 1200~keV, (c) 1200 to 1800~keV, and (d) 1800 to 2400~keV. All other excited states have been observed to possess a decay branch through this level.}
\end{figure}

A decay scheme was primarily built upon the 181-keV $E2$ transition connecting the 2$^+_1$ and 0$^+_1$ levels. Almost all other excited states that were identified cascade through this 2$^+_1$ level and, as such, their associated $\gamma$~rays are found to be in coincidence with the strong 181-keV $\gamma$~ray. The one exception to this is the 1657-keV level; $\gamma$ transitions from this level to the ground state and 739-keV level were observed, but there was no evidence of a transition directly to the 2$^+_1$ state. Twelve levels have decay branches that feed into the 514-, 739-, or 822-keV levels, which then proceed to decay to the 181-keV level or ground state. The remaining 19 levels were observed to only have decay branches to the 181-keV level, and in some cases directly to the ground state. The background-subtracted projection of the $\beta$-correlated $\gamma-\gamma$ matrix, gated on the 181-keV transition, is presented in Fig. \ref{Figure3}. This was used as the starting point in identifying which $\gamma$~rays belong to transitions in \nuc{146}{Ba}. There is a small contribution from random coincidences with strong $\gamma$~rays of \nuc{146}{Ce} and \nuc{146}{Nd}, 2$^+_1$$\rightarrow$~0$^+_{\rm g.s.}$ transitions (258~keV and 454~keV). 

\begin{figure}[h!]
\includegraphics[width=8.5cm]{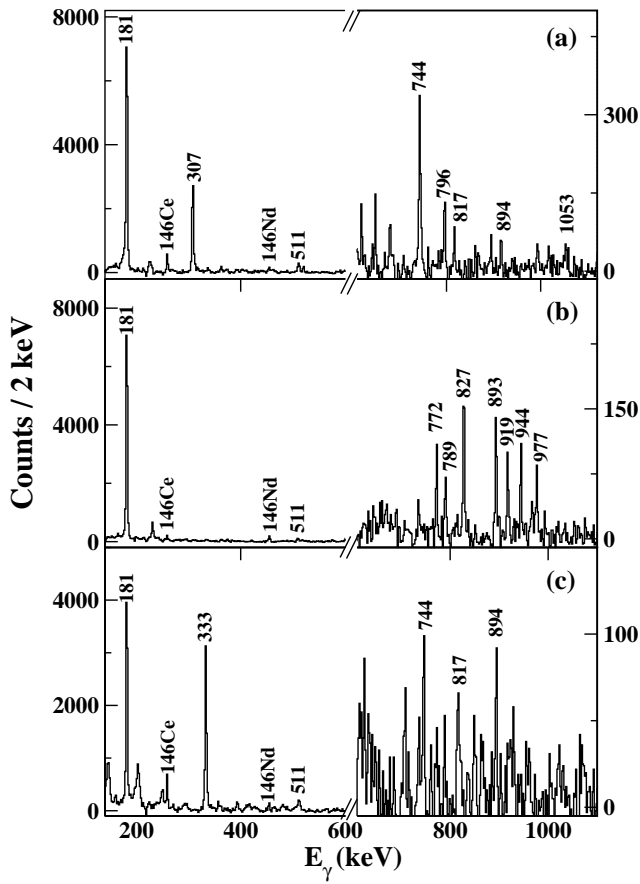}
\caption{\label{Figure4} Background-subtracted projections of the $\beta$-$\gamma$-$\gamma$ coincidence histogram gated on (a) the 333-keV, 4$^+_1$$\rightarrow$~2$^+_1$ transition, (b) the 558-keV, 1$^-_1$$\rightarrow$~2$^+_1$ transition, and (c) the 307-keV, 3$^-_1$$\rightarrow$~2$^+_1$ transition. Gamma rays from \nuc{146}{Ba} are labelled by their energies. Random coincidence events from \nuc{146}{Ce} and \nuc{146}{Nd} 2$^+_1$$\rightarrow$~0$^+_1$ transitions are labelled.}
\end{figure}

Placement of $\gamma$~rays in the level scheme was confirmed by gating on $\gamma$~rays above and below the known 4$^+_1$, 1$^{-}_1$, and 3$^{-}_1$ levels. The 4$^+_1$ state decays via a 333-keV $E2$ transition to the 2$^+_1$ level, whereas the 1$^{-}_1$ (558~keV and 739~keV) and 3$^{-}_1$ (307~keV and 640~keV) states each have two decay paths. Figure \ref{Figure4} provides the background-subtracted matrix projections with an appropriate gate for each of these levels.  
 
%%%%%%%%%%%%%%%%%%%%%
\subsection{The decay scheme}\label{subsec:decay scheme}
%%%%%%%%%%%%%%%%%%%%%

\begin{figure*}[h!!]
\includegraphics[width=15cm]{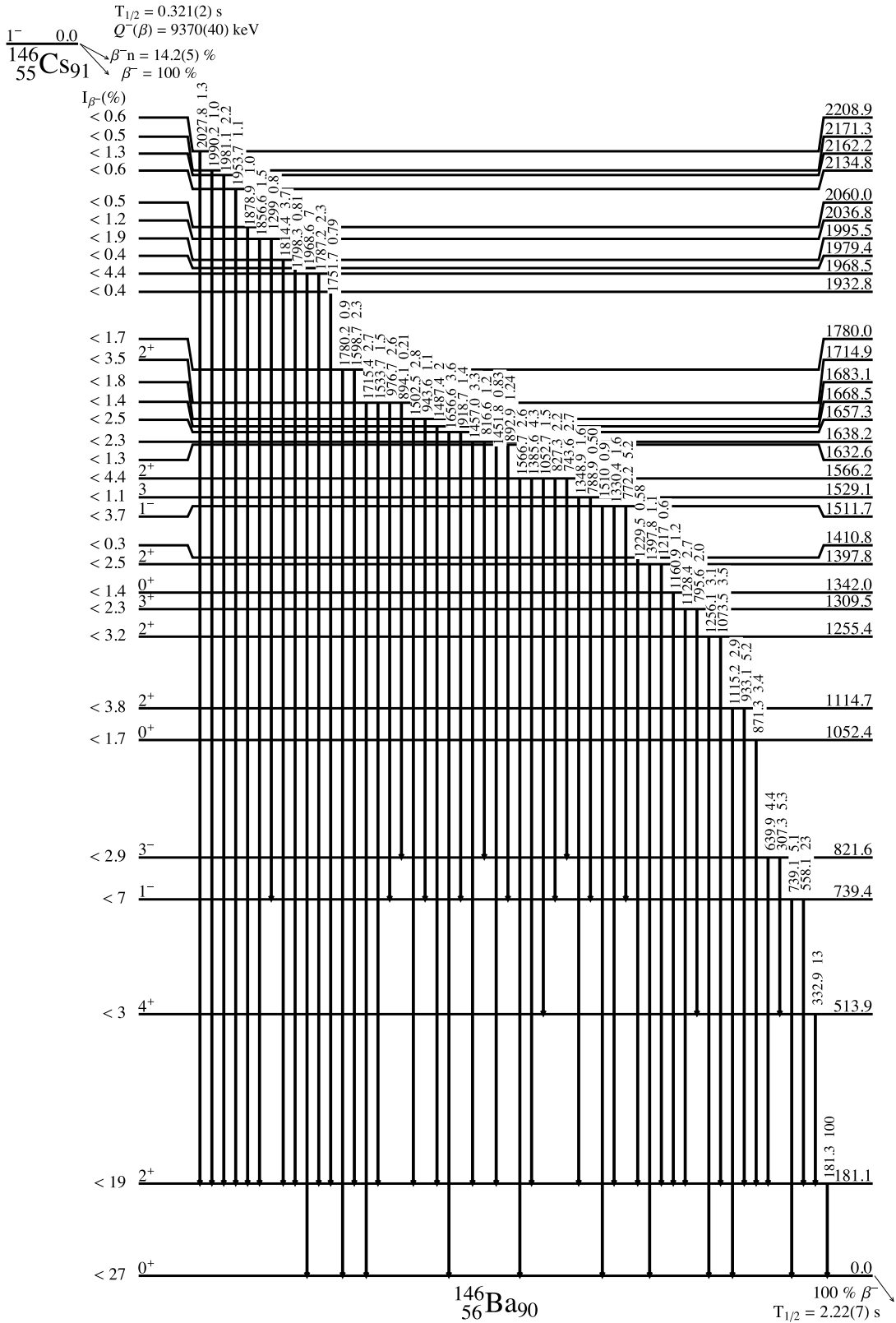}
\caption{\label{Figure5}Decay scheme of \nuc{146}{Ba} populated in \nuc{146}{Cs} $\beta^-$ decay. In total, 31 excited states with 54 $\gamma$-ray transitions have been identified. Labels indicate the energy and relative intensity of each transition. For absolute intensity per 100 decays, multiply by 0.42(5). I$_{\beta^-}$ values were determined by an intensity balance between the $\gamma$~rays feeding and de-exciting each level, as discussed in the text.}
\end{figure*}

The Nuclear Data Sheets list nine confirmed levels, two tentative excited states, and 21 $\gamma$-ray transitions for \nuc{146}{Ba} from previous $\beta$-decay studies \cite{Reference30}. In this work, we report a total of 31 excited states with 54 $\gamma$-ray transitions, offering a significant increase in the known \nuc{146}{Ba} level structure. Our proposed expansion of the known decay scheme is displayed in Fig. \ref{Figure5}. 

These data confirm the correct placement of 21 $\gamma$-ray transitions between the known levels. Thirty-two $\gamma$~rays are listed in Ref. \cite{Reference30} without placement in the adopted decay scheme. This work has identified 19 of these, which were subsequently placed in the new scheme. Thirteen remaining $\gamma$~rays in the adopted list of Ref. \cite{Reference30} have not been observed, suggesting that, in fact, they are not associated with \nuc{146}{Ba}. Furthermore, we have identified 14 new $\gamma$~rays. Upper limits have been applied to an additional 17 unobserved $\gamma$ transitions, two of which are listed in the ENSDF adopted list of $\gamma$~rays for \nuc{146}{Ba}.

While the level scheme has been extended extensively from what was previously known, the highest level observed lies at $\sim$2.2~MeV, i.e., $\sim$3~MeV below the neutron separation energy. It is possible that direct $\beta$ feeding to weak states within this energy range occurs which is not measurable with discrete-line spectroscopy. Resolving this issue will require some other technique, such as Total Absorption Gamma-ray Spectroscopy. 

%%%%%%%%%%%%%%%%%%%%%
\subsection{$\gamma$-ray intensities}\label{subsec:intensities}
%%%%%%%%%%%%%%%%%%%%%

The spins of a few low-lying states have been tentatively assigned in the literature. Where possible, these assignments have been used in calculating the conversion coefficient for each transition with the BrICC code \cite{Reference31}. The conversion coefficient for the 181-keV transition is 0.241(4). As the remainder of observed transitions are greater than 300~keV, conversion coefficients are expected to be negligible. 

Relative $\gamma$-ray intensities have been determined from the observed number of counts in the $\beta - \gamma$ singles spectrum, corrected for $\gamma$- and $\beta$-detection efficiency, such that the 2$^+_1$$\rightarrow$~0$^+_{1}$ is normalized as 100. The absolute $\gamma$-ray normalization was calculated accounting for the known $\beta$-delayed neutron emission of \nuc{146}{Cs} (14.2$\%$ \cite{Reference30}), and assuming no delayed neutron emission from \nuc{146}{Ba}. This was achieved from a comparison of the 181-keV \nuc{146}{Ba} $\gamma$~ray to the 141-keV $2^-_3$$\rightarrow$~2$^-_1$ transition in the daughter, \nuc{146}{La} \cite{Reference30}. In this procedure, it is assumed that the contribution of \nuc{146}{Ba} in the beam was negligible as these ions would be extracted from the gas catcher in a 2$^+$ charge state, wheras a 1$^+$ \nuc{146}{Cs} beam was selected through the separator. Any $\gamma$ decay in \nuc{146}{La} results from a $\beta$~decay of \nuc{146}{Ba} in the 0$^+$ ground state. A low-energy, high-spin (6$^-$) isomer is reported in \nuc{146}{La} \cite{Reference32}. It is assumed that this isomer is not populated and the 141-keV transition has an absolute intensity $I_{\gamma}$~=~20.2(20)$\%$ \cite{Reference30}. The number of efficiency-corrected counts observed in both peaks is given in Table \ref{Table1}. Using the adopted value of $I_{\gamma}$ for the 141-keV \nuc{146}{La} $\gamma$~ray and the $\beta$-delayed neutron branch, the `total' number of parent \nuc{146}{Cs} decays was determined. The ratio of the 181-keV $\gamma$-ray intensity to this parent population gives the ``normalization'' for Table \ref{Table1} as 0.42(5), i.e., there are 42 181-keV $\gamma$~rays per 100 \nuc{146}{Cs} decays. 

The intensity balance also allows an estimate of the $\beta$-branch of \nuc{146}{Cs} to the ground state of \nuc{146}{Ba}. Even after correcting for internal conversion, the total identified decay to the ground state is less than the population of \nuc{146}{La}, so we infer the ground-state feeding in \nuc{146}{Ba} to be $\textless$~27$\%$. This is only an estimate, as any extra unobserved feeding to the ground state from high-lying states in \nuc{146}{Ba} will reduce this number. We have examined the distribution of $\beta$-feeding to the states we have observed. This was estimated by studying the intensity balance of $\gamma$~rays populating and depopulating each level. This approach is limited by the completeness of the level scheme; if low-intensity transitions from high-lying states are missed, then this will distort the inferred feeding. An indication of the level of ``missing'' $\gamma$-ray strength can be seen through the $\sim$2$\%$ population of low-lying 4$^+$ and 3$^-$ states which are forbidden decays, so should receive very little direct $\beta$-population. Thus, the $\beta$-feeding intensities in Fig. \ref{Figure5} are shown as upper limits. The key observation is that the feeding pattern is very widely distributed and no individual high-lying state is strongly populated. Clearly, there is little overlap between the wave function of the \nuc{146}{Cs} ground state and any of the excited levels in \nuc{146}{Ba}.

A summary of the data, including $\gamma$-ray energies and intensities is provided in Table \ref{Table1}. For some levels that were identified from $\gamma - \gamma$ coincidence data, the corresponding transition to the ground state was not observed in the singles data. An upper limit on the relative intensity of such transitions has been determined using the intensity of the weakest $\gamma$~ray that was observable. These have not been included in intensity balances or normalization. 

%%%%%%%%%%%%%%%%%%%%%
\subsection{\textit{J}$^{\pi}$ assignments}\label{subsec:spin assignments}
%%%%%%%%%%%%%%%%%%%%%

Spin values can be constrained for many observed \nuc{146}{Ba} excited states from detailed inspection of $\gamma$-ray transitions to levels with firm spin-parity assignments and $\beta$-decay selection rules. The data were not sufficient to confirm these assignments through $\gamma-\gamma$ directional correlation measurements. The $J^{\pi}$~=~1$^-$ spin and parity of the parent is well known, having been measured via high-resolution laser spectroscopy \cite{Reference33}. 

It is expected that the observed levels in \nuc{146}{Ba} are mostly populated via allowed (1$^-$$\rightarrow$~0$^-$, 1$^-$, 2$^-$) or first-forbidden decays (1$^-$$\rightarrow$~0$^+$, 1$^+$, 2$^+$, 3$^+$). Observation (or non-observation) of $\gamma$ transitions to the 0$^+$ ground state can be used to further constrain the spin assignment. Upper limits for relative intensities of unobserved $\gamma$ transitions have been discussed above. The yrast levels lying below 1-MeV excitation have been reported in angular-correlation measurements \cite{Reference27}. The non-yrast states above 1 MeV typically decay via low-multiplicity cascades through the 2$_1^+$ level. States of $J = 1$ or $J = 2$ are also seen to decay directly to the ground state. 

\begin{itemize}
\item
{\it The 1115- and 1255-keV levels} \newline
Excited states at 1115 keV and 1255 keV both $\gamma$ decay to the 2$_1^+$ and 0$_{1}^+$ levels, therefore $J = 1, 2$ assignments are possible. No transitions to the negative-parity states were observed, suggesting that these are positive-parity states. The IBA-1 calculations (discussed below) also predict that the 2$_2^+$ lies at 1101 keV. We assign the 1115-keV and 1255-keV levels to be the 2$_2^+$ and 2$^+_3$ level, respectively.
 \item
{\it The 1310-keV level} \newline
The 1310-keV state feeds the 4$_1^+$ and 2$_1^+$ levels with no observed direct feeding to the 0$_{1}^+$ level. Given that the $\gamma$ transitions only involve positive-parity states, we suggest this is the 3$_1^+$ level. 
\item
{\it The 1342-keV level} \newline
Since the only $\gamma$ transition from the 1342-keV level is to the 2$_1^+$ state, we assign this to be the 0$_2^+$ level. The $\beta$-feeding is large enough that, if this were not a 0$^+$ state, $\gamma$ transitions to other levels would be expected to have intensities above the upper limit for non-observation. 
\item
{\it The 1398-keV level} \newline
This excited state exhibits $\gamma$-decay characteristics similar to those of the 1115- and 1255-keV levels, therefore a 2$^+$ spin-parity assignment is appropriate. 
\item
{\it The 1512-keV level} \newline
We propose a 1$^-$ assignment to this state since it exhibits strong feeding to other low-spin $(J = 0, 1, 2)$ states of both positive and negative parity, with an enhanced branch to the 1$_1^-$ state.
\item
{\it The 1529-keV level} \newline
This level decays to the 1$_1^-$ and 2$_1^+$ levels, with no observed direct feeding to the ground state. A 2$_1^-$ assignment is allowed, however the strong branch to the 2$_1^+$ level favors a spin assignment of $J = 3$. 
\item
{\it The 1566- and 1715-keV levels} \newline
Strong $\beta$ feeding and subsequent $\gamma$ decays to all low-lying yrast states imply a uniquely constrained $J^{\pi}$ = 2$^+$ spin-parity for these levels. \\[-0.2cm]
\end{itemize}

For the remaining states, it has not been possible to draw any solid conclusion pertaining to their appropriate spin-parity assignments. In a few cases, a higher spin assignment is favored since no decay to the ground state was observed. However, the $\beta$ feeding is weak and so it was not possible to ascertain whether the $\gamma$ transition does not exist, or lies below the observation limit of the data.

%%%%%%%%%%%%%%%%%%%%%%%%%%%%%%%%%%%%%%%%%
%%%% Discussion
%%%%%%%%%%%%%%%%%%%%%%%%%%%%%%%%%%%%%%%%%

\section{Discussion}\label{sec:discussion}

With strong octupole correlations prevalent in this region, double-octupole vibrations may be observable. The excitation signature of this collective mode would be a two-phonon multiplet (0$^+$, 2$^+$, 4$^+$, 6$^+$) located at approximately twice the excitation energy of the 3$_1^-$ state. The 0$^+$ and 6$^+$ members decay via two $E3$ transitions to the 3$^-$ level and then the ground state, and the 2$^+$ and 4$^+$ members decay via enhanced $E1$ transitions. While the 4$^+$ and 6$^+$ members will not be populated in $\beta$~decay, one might expect to find the 0$^+$ and 2$^+$ members at $\sim$(2$\times$822) keV. The 1638-keV level feeds the 3$_1^-$ and 2$_1^+$ states, and does not $\gamma$ decay to the ground state; therefore, a case can be made that this state corresponds to the 0$^+$ member of the two-octupole phonon multiplet. Similarly, the 1715-keV level also decays through the 3$_1^-$ state and may possibly be associated with the 2$^+$ member of this multiplet. However, additional data are required to draw firm conclusions about the observation of double-octupole vibrations in \nuc{146}{Ba}. 

Key spectroscopic observables which differentiate between models describing nuclear shape changes are the excitation energies, spins and parities of low-lying non-yrast states, particularly the lowest few $J^{\pi} = 0^+$ and $J^{\pi} = 2^+$ levels, their electromagnetic decay properties, and evidence for collective bands built upon them. Understanding the development of collective behavior at the beginning of the rare-earth region has evolved with our capacity to constrain these observables. The focus of this work is on determining spins and parities of these important levels in \nuc{146}{Ba}. In this respect, the project was only partially successful. The present data set has revealed many new, higher-lying states which do not inform this particular aspect. The data were insufficient for $\gamma-\gamma$ directional correlation measurements. However, the enhanced sensitivity does offer the opportunity for observation of some new low-intensity decays between key low-lying states which constrain their possible spins, sometimes uniquely. There is strong evidence that the 181-, 1115-, and 1255-keV levels are the $J^{\pi} = 2^+_1, 2^+_2$, and $2^+_3$ states. We use these assignments in the following discussion. The remaining uncertainty with these assignments is the observation of several other low-lying states which are interspersed between these levels and for which a firm spin assignment could not be made. As such, the possibility that these are additional $J^{\pi} = 2^+$ levels cannot be ruled out.

The decay scheme of \nuc{146}{Ba} was investigated within the framework of the Interacting Boson Approximation (IBA) \cite{Reference34} by Scott {\it et al}., \cite{Reference27} and, more recently, by Gupta and Saxena \cite{Reference35}. Both of these studies used a $\chi$-parameter of  $\chi = - \sqrt7/2$, which corresponds to an axially symmetric potential in the $\gamma$ degree of freedom centered at $\gamma = 0^{\circ}$. A general study of the $N = 90$ transition region in the IBA \cite{Reference36} indicates that this is unlikely to be the case. 

\begin{figure}[h!]
\includegraphics[width=8.5cm]{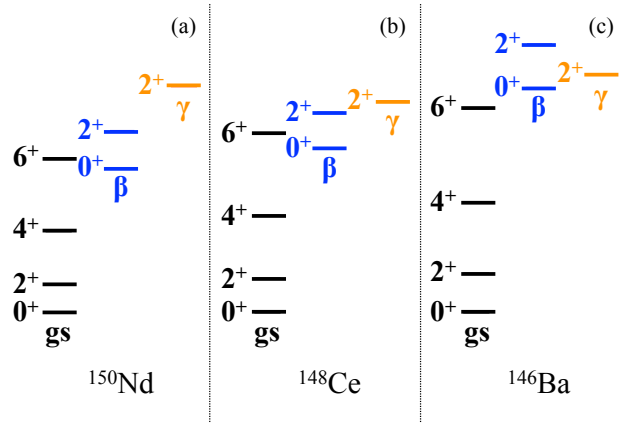}
\caption{\label{Figure6} [Color online] Truncated level schemes showing the lowest-lying members of the ground-state, $\beta-$ and $\gamma-$vibrational bands in (a) \nuc{150}{Nd}, (b) \nuc{148}{Ce}, and (c) \nuc{146}{Ba}. The individual band sequences are labelled for each $N = 90$ isotope.}
\end{figure}

\begin{center}
\begin{figure}[h!]
\includegraphics[width=8.5cm]{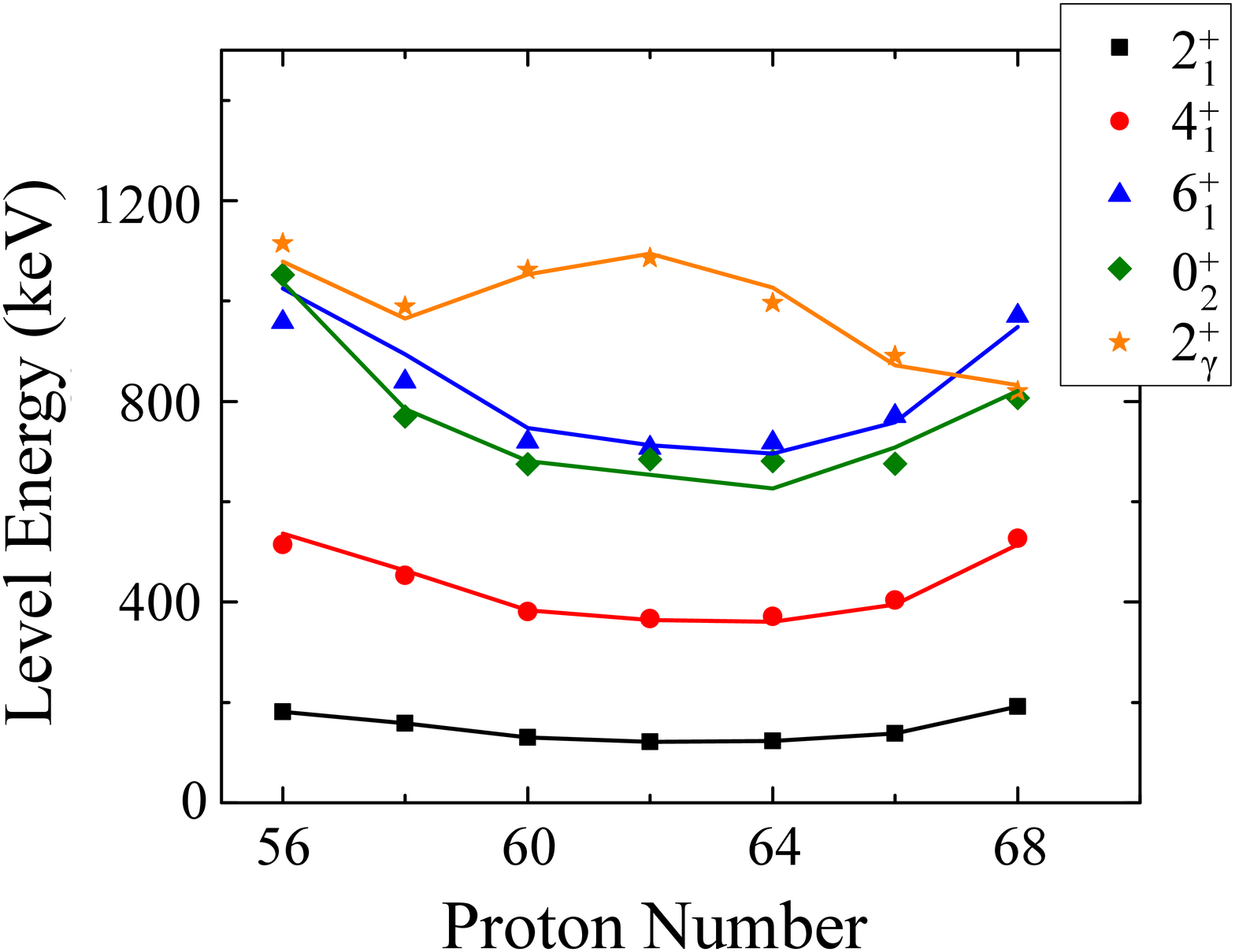}
\caption{\label{Figure7} [Color online] IBA fits (lines) from this work to experimental data (symbols) for the $N = 90$ isotones.}
\end{figure}
\end{center}

Truncated level schemes of the lowest members of the ground-state, $\beta$- and $\gamma$-vibrational bands for \nuc{150}{Nd} \cite{Reference6}, \nuc{148}{Ce} \cite{Reference37}, and \nuc{146}{Ba} (this work) are presented in Fig. \ref{Figure6}. The key signature of non-axial behavior in \nuc{146}{Ba} lies in the location of the $2^+_2$ state at 1115 keV with respect to the $0^+_2$ and $2^+_3$ levels. Indeed, in line with the systematics emerging from Fig. \ref{Figure6} and Fig. \ref{Figure7}, we interpret the $2^+_2$ level as the bandhead of the $\gamma$-vibrational sequence and associate the $0^+_2$ and $2^+_3$ with the (quasi-$\beta$) band. Hence, the $\gamma$ and $\beta$ excitations lie remarkably close in energy, to the extent that the ``$2^+_{\beta}$'' and ``$2^+_{\gamma}$'' locations are reversed with respect to the heavier isotones. However, these assignments, and the association of a projection of angular momentum on the axis of deformation, $K$, are only rigorously applicable for axially symmetric nuclei. In fact, in any non-axially symmetric case, these states mix, especially in a case like this where the moments of inertia would suggest that their unperturbed positions are nearly degenerate. The relative lowering of the excitation energies of the $J^{\pi} = 2^+_{\gamma}$ levels in \nuc{146}{Ba} and \nuc{148}{Ce}, shown by the ratio E($2^+_{\gamma}$)/E($2^+_1$) = 6.2 and 6.3, respectively, can be compared with 8.2 in \nuc{150}{Nd} and 8.9 in \nuc{152}{Sm}. This is an indication that the triaxial potential energy is soft for \nuc{146}{Ba} and \nuc{148}{Ce}. Such an observation tends to disfavor any interpretation in terms of the X(5) geometric model, which is based on a stiff, axially-symmetric potential in the $\gamma$ degree of freedom. 

In an effort to better understand this evolution of structure in barium, and indeed along $N = 90$, IBA calculations were performed. The simplest version of the model was used, which makes no distinction between proton and neutron bosons (IBA-1), and employed the Extended Consistent-Q Formalism (ECQF) \cite{Reference38}. The entire IBA space can be described with a two-parameter Hamiltonian incorporating a term related to the $\beta$ deformation, $\zeta$, and one associated with the degree of axial asymmetry, $\chi$. The IBA-1 Hamiltonian is given by \cite{Reference39,Reference40}:

\begin{equation}
H_{\rm{IBA-1}}(\zeta) = c\left[ (1-\zeta)\hat{n}_d - \frac{\zeta}{4N_B} \hat{Q}^{\chi} \cdot \hat{Q}^{\chi} \right] , \\
\end{equation}

\noindent
where

\begin{equation}
\hat{Q}^{\chi} = (s^{\dagger}\tilde{d} + d^{\dagger}s) + \chi(d^{\dagger}\tilde{d})^{(2)}, \\
\end{equation}

\noindent
and $ \hat{n}_d = d^{\dagger}\cdot\tilde{d} $. The parameters for the fits are included in Table \ref{Table2}. A comparison between the experimental and calculated low-lying level energies is given in Fig. \ref{Figure7}. The calculations are in excellent agreement with the data, agreeing usually at the 10$\%$ level or better, with the best fit for \nuc{146}{Ba} corresponding to a $\gamma$-soft shape.

\begin{table}[b!]
\caption{\label{Table2} Parameters $\zeta$ and $\chi$ used for each $N = 90$ isotone in the IBA fits of this work.}
\begin{ruledtabular}
\begin{tabular}{cdd}
 \text{Isotone} & 
 \text{$\zeta$} & 
 \text{$\chi$} \\
 \hline
 
Ba	& 0.732	& -0.78	\\
Ce	& 0.653	& -0.95	\\
Nd	& 0.632	& -1.03	\\
Sm	& 0.597	& -1.21	\\
Gd	& 0.595	& -1.10	\\
Dy	& 0.615	& -0.85	\\
Er	& 0.633	& -0.61	\\

\end{tabular}
\end{ruledtabular}
\end{table}

\begin{center}
\begin{figure}[h!]
\includegraphics[width=7.0cm]{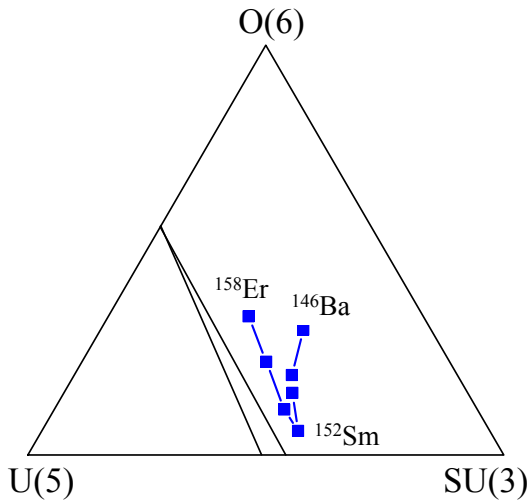}
\caption{\label{Figure8} [Color online] Trajectories within the IBA symmetry triangle for the $N = 90$ isotonic chain, mapped according to the polar coordinate system of \cite{Reference36}. The slanting lines enclose the region of phase coexistence and phase transition.}
\end{figure}
\end{center}

Figure \ref{Figure8} highlights the evolution within the so-called `Casten triangle' \cite{Reference41} of the $N = 90$ isotones from \nuc{146}{Ba} to \nuc{158}{Er}. Only with $Z = 62, 64$ (Sm and Gd) are the $N = 90$ isotones near the axial ($\chi \sim-1.32$) route from U(5) to SU(3). Both heavier and lighter isotones are best fitted with parameters deep in the interior of the triangle, i.e., they follow the trend to deformation along loci corresponding to non-axial shapes. Both above and below $Z = 64$, the trend of non-axial behavior seems to be quite symmetric.

The near degeneracy of the $J^{\pi} = 2^+_2$ and 0$^+_2$ levels is quite rare and has been discussed as a possible signature for nuclei with properties lying along the so-called ``Alhassid-Whelan Arc of Regularity \cite{Reference42}''; e.g., a small number of nuclei which have statistically regular spectra that are found in the mainly chaotic IBA parameter space. An experimental signature of nuclei which may exhibit this regular behavior has been defined as those having $|E(2^+_2)-E(0^+_2)|/E(2^+_2) \leq 0.025$ \cite{Reference43}. In \nuc{146}{Ba}, this quantity is small, 0.055, but just outside the prediction for identifying nuclei on the non-chaotic arc. However, this simple experimental signature does not always exactly follow the trajectory of ``regular'' nuclei which are inferred from a full statistical analysis of the spectra \cite{Reference44}. Interestingly, both $N = 90$ \nuc{156}{Dy} and \nuc{158}{Er} \cite{Reference45} have been previously identified as nuclei lying close to the regular region \cite{Reference43}. The fact that \nuc{146}{Ba} exhibits a similar degeneracy appears related to $\gamma$ softness and the symmetry of these shapes above and below axially symmetric $Z = 62$, \nuc{152}{Sm}.

%%%%%%%%%%%%%%%%%%%%%%%%%%%%%%%%%%%%%%%%%
%%%% Conclusions
%%%%%%%%%%%%%%%%%%%%%%%%%%%%%%%%%%%%%%%%%

\section{CONCLUSIONS}\label{sec:conclusions}

A detailed $\beta$-decay spectroscopy measurement has been conducted on the neutron-rich exotic nucleus \nuc{146}{Ba}. This represents the first results from the recently-commissioned decay-spectroscopy station for low-energy CARIBU beams at Argonne National Laboratory. The experimental arrangement had a high sensitivity to weak $\gamma$-ray transitions and, hence, enabled the study of excited states not strongly populated via $\beta$~decay. Inspection of these low-intensity transitions has allowed spin constraints for low-lying levels, which have also been considered within the IBA framework. The $N = 90$ isotones are situated close to, but slightly to the right, of the phase-transitional region predicted by the IBA. They follow a symmetric behavior about \nuc{152}{Sm} ($Z = 62$) which exhibits the highest degree of axial symmetry. Moving away from \nuc{152}{Sm}, isotones of both larger and smaller $Z$ appear to exhibit increasing $\gamma$ softness.  \\

%%%%%%%%%%%%%%%%%%%%%%%%%%%%%%%%%%%%%%%%%
%%%% Acknowledgements
%%%%%%%%%%%%%%%%%%%%%%%%%%%%%%%%%%%%%%%%%

\section{ACKNOWLEDGEMENTS}\label{acknowledgements}

The authors wish to acknowledge the Physics Support group at Argonne National Laboratory and engineering work of the Submillimeter-Wave Technology Laboratory, University of Massachusetts Lowell. Figure \ref{Figure5} in this article has been created using the LevelScheme scientific figure preparation system [M. A. Caprio, Comput. Phys. Commun. 171, 107 (2005), \url{http://scidraw.nd.edu/levelscheme}]. This material is based upon work supported by the U.S. Department of Energy, Office of Science, Office of Nuclear Physics under Grant Nos. DE-FG02-94ER40848 and DE-FG02-94ER40834, and Contract No. DE-AC02-06CH11357, and the U.S. National Science Foundation under Grant No. 1064819. This research used resources of ANL's ATLAS facility, which is a DOE Office of Science User Facility.\\

%%%%%%%%%%%%%%%%%%%%%%%%%%%%%%%%%%%%%%%%%
%%%% Bibliography
%%%%%%%%%%%%%%%%%%%%%%%%%%%%%%%%%%%%%%%%%


%%%%%%%%%%%%%%%%%%%%%%%%%%%%%%%%%%%%%%%%%
%%%% End of text
%%%%%%%%%%%%%%%%%%%%%%%%%%%%%%%%%%%%%%%%%

\end{document}